# Semiclassical spin-bath calculation of the nitrogen-isotope effect on $NV^-$ ensemble coherence in diamond

Chikara Shinei[1,2,*]

[1]Department of Applied Physics, Institute of Pure and Applied Sciences, University of Tsukuba, Tsukuba, Ibaraki 305-8573, Japan

[2]Japanese-French Laboratory for Semiconductor Physics and Technology J-FAST, CNRS, University Grenoble Alpes, Grenoble INP, University of Tsukuba, Tsukuba, Japan

*shinei.chikara.fb@u.tsukuba.ac.jp

The ratio $T_2/T_2^*$ of the Hahn-echo and Ramsey times of nitrogen-vacancy ($NV^-$) ensembles is independent of the nitrogen concentration [N] and is ≈ 16 in diamond of natural isotopic abundance ($^{14}N$), yet a recent $^{15}N$-doped ensemble magnetometer gives only 8.05 ± 0.18 in the single-quantum convention. We ask whether the nitrogen nuclear isotope alone can cause such a reduction. Using a semiclassical spin-bath model with the Jahn-Teller-resolved P1 hyperfine tensor, we show that the fraction of P1 pairs that are hyperfine-degenerate, and hence free to flip-flop, is 1/4 for $^{14}N$ ($I = 1$) but 5/16 for $^{15}N$ ($I = 1/2$). Simulations over [N] = 0.1–100 ppm confirm that this shortens $T_2$ while leaving $T_2^*$ exactly unchanged: $T_2(^{14}N)/T_2(^{15}N) = 1.118 \pm 0.011$, independent of [N] and 11σ above unity. The measured contrast, 2.07 ± 0.25, is 3.8σ larger, identifying resonant-channel counting as a real but partial contribution and setting a quantitative benchmark for many-body calculations.

Ensembles of negatively charged nitrogen-vacancy ($NV^-$) centers in diamond underpin a broad class of solid-state magnetometers [1]. Their sensitivity is governed by the ensemble dephasing time $T_2^*$ for broadband dc sensing and by the Hahn-echo coherence time $T_2$ for narrowband ac sensing. In nitrogen-rich diamond both times are limited by the bath of substitutional nitrogen (P1) electron spins, and both scale inverse-linearly with the nitrogen concentration [N] [2].

Which nuclear species surround a spin qubit, and in what isotopic proportion, is increasingly treated as a design variable. Cluster-correlation-expansion (CCE) simulations [3] of more than 12 000 host compounds show that each nuclear species acts as an independent bath whose contribution to $T_2$ scales inversely with its density and is fixed by its gyromagnetic ratio and nuclear spin I [4]. In NV ensembles, however, the dominant bath is formed by the electron spins of the P1 centers, each carrying its own nitrogen nucleus, $^{14}N$ ($I = 1$) or $^{15}N$ ($I = 1/2$). Does the isotope of that attached nucleus change how fast the electron bath fluctuates, and with it the coherence of the NV?

That the nitrogen nucleus matters was shown by Park et al., who combined CCE with first-principles hyperfine tensors [5]. The $^{14}N$ hyperfine interaction shifts each P1 transition by tens of megahertz, by an amount that depends on the nuclear state $m_I$ and on the Jahn-Teller axis of the center [6,7], and two P1 centers can exchange spin only if these shifts coincide; switching the hyperfine interaction off shortens the calculated $T_2$ by a factor of about four [5]. Because the suppression is a question of how many hyperfine levels a pair can match, the nuclear multiplicity $2I + 1$ is a natural control parameter: $^{15}N$ offers two levels where $^{14}N$ offers three.

The question is not confined to dense ensembles. The 6.8 ms Hahn-echo time recently reported for a single NV in $^{13}C$-depleted diamond is again limited by a P1 bath, at a nitrogen content of order 10 ppb [8], and the hyperfine part of the resonance condition for a P1 flip-flop is a property of each center alone, independent of how far apart the centers are.

Because $T_2$ and $T_2^*$ share the same 1/[N] scaling, their ratio is independent of [N], whose absolute value is typically uncertain by 50 %. Bauch et al. found $T_2/T_2^* \approx 16$ across 25 diamonds of natural isotopic abundance (99.6 % $^{14}N$) [2]. Barry et al. recently reported an ensemble magnetometer built on a deliberately $^{15}N$-doped layer, chosen to reduce the number of required microwave frequencies [9]. In the double-quantum (DQ) basis [10] they measure $T_2^{*,DQ}$ = 14.0(1) μs and $T_{2,DQ}$ = 142(3) μs [p = 1.14(3)]. Converting to the single-quantum (SQ) convention of Ref. [2] — a factor of two for $T_2^*$ and, since the sample is in the quasi-static regime, $2^{2/3} = 1.59$ for $T_2$ (Supplementary Material, Sec. S4) — gives $T_2/T_2^* = 8.05 \pm 0.18$, a factor 1.99 below the $^{14}N$ value.

Can the nitrogen isotope, by itself, account for this contrast? The expectation is asymmetric. $T_2^*$ is set by the quasi-static dipolar field of the P1 bath at the NV and is blind to the nuclear spin, whereas $T_2$ also depends on the bath correlation time $\tau_c$, which is governed by P1–P1 flip-flops and hence by the number of hyperfine levels. We test this with the semiclassical spin-bath model of Bauch et al. [2], extended with the Jahn-Teller-resolved hyperfine tensor [5]. The calculation is intentionally not many-body: we ask whether a transparent, parameter-free counting argument reproduces the direction and part of the magnitude of the contrast, as a benchmark for a future CCE treatment [3,11].

We follow Ref. [2] throughout. An $NV^-$ center is placed at the origin of a diamond lattice and P1 centers are assigned to random substitutional sites with probability [N]. Under the secular approximation the dipolar interaction between spins i and j is

$$H_{ij} = C_\parallel^{ij} S_z^i S_z^j + C_\perp^{ij}\left(S_+^i S_-^j + S_-^i S_+^j\right), \tag{1}$$

with, in a frame whose z axis lies along the [111] crystal axis (the NV symmetry axis and the applied field direction),

$$C_\parallel = \frac{\mu_0}{4\pi}\frac{\hbar^2\mu_e^2}{r^3}\left(1 - 2\cos^2\theta\right),$$
$$C_\perp = \frac{\mu_0}{4\pi}\frac{\hbar^2\mu_e^2}{2r^3}\left(1 - \frac{1}{4}\sin^2\theta\right). \tag{2}$$

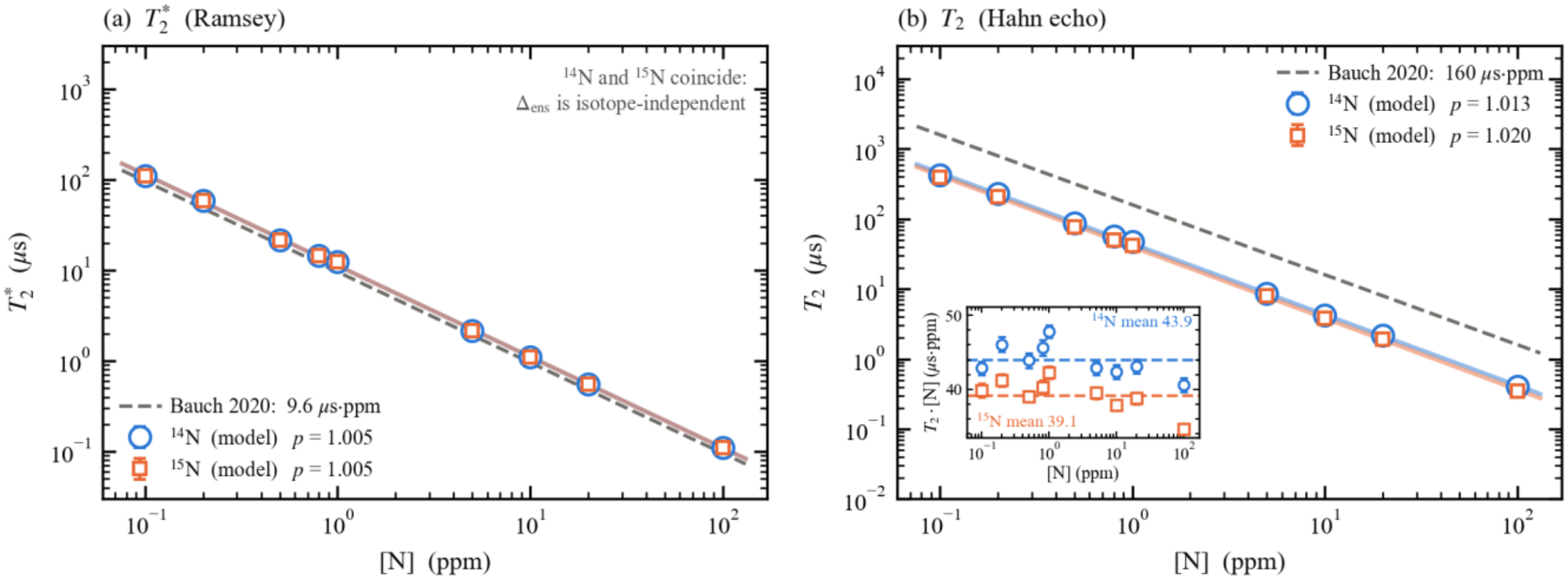

FIG. 1. Simulated ensemble coherence times versus nitrogen concentration for $^{14}N$ (blue circles) and $^{15}N$ (orange squares) P1 baths, with 1σ uncertainties propagated from the fits of Eqs. (7) and (8). Solid lines are weighted power-law fits; dashed grey lines are the measured scalings of Ref. [2], 9.6 μs·ppm for $T_2^*$ and 160 μs·ppm for $T_2$. (a) $T_2^*$ is identical for the two isotopes, as required by Eq. (3). (b) $T_2$ is shorter for $^{15}N$ at every concentration; the model underestimates its absolute value by a factor of about 3.6. Inset: $T_2$·[N], which removes the common 1/[N] dependence; dashed lines are the weighted means, and $^{14}N$ lies above $^{15}N$ at every concentration.

The NV zero-field splitting suppresses NV–P1 flip-flops, so the NV experiences the P1 bath as an effective longitudinal field of root-mean-square strength

$$\Delta_{\text{single}}^2 = \sum_i \left(\frac{C_\parallel^{\text{NV},i}}{2}\right)^2. \tag{3}$$

Each P1 center possesses a Jahn-Teller (JT) distortion axis along one of the four ⟨111⟩ directions, chosen at random and static on the timescale of the experiment [5–7]. The hyperfine tensor is axially symmetric about the JT axis with principal values $A_\parallel$ = 114.03 MHz and $A_\perp$ = 81.31 MHz for $^{14}N$. With the magnetic field along one particular [111] axis, one JT orientation (probability 1/4) is parallel to the field while the other three (probability 3/4) make an angle of 109.47° with it. The secular hyperfine splitting is therefore neither $A_\parallel$ nor $A_\perp$ but

$$A_{\text{eff}}(\theta) = \sqrt{A_\parallel^2\cos^2\theta + A_\perp^2\sin^2\theta}, \tag{4}$$

giving $A_{eff}$ = 114.03 MHz for the on-axis orientation and $A_{eff}$ = 85.57 MHz for the three off-axis orientations. The difference, 28.5 MHz, reproduces the "$A_\parallel - A'_\parallel \approx 29$ MHz" level shift identified in the CCE analysis of Ref. [5], and provides an independent check of Eq. (4). For $^{15}N$ the principal values are scaled by $\gamma(^{15}N)/\gamma(^{14}N) = -1.4028$, giving $A_{eff}$ = 159.95 MHz and 120.03 MHz respectively.

The on-site hyperfine energy of a P1 center is $E = A_{eff} \cdot m_I$, with $m_I$ drawn uniformly from {−1, 0, +1} for $^{14}N$ and from {−1/2, +1/2} for $^{15}N$, since all nuclear states are equally populated at room temperature.

A flip-flop between two P1 spins is a two-level problem with coupling $\Omega = C_\perp(ij)$ and detuning δ equal to the difference of the local fields experienced by the two spins, including the hyperfine (Overhauser) contribution and the fields of neighbouring P1 centers. In the presence of bath dephasing at rate $\Gamma_d$, the population transfer rate is

$$R_{\text{flip}} = \frac{\Omega^2}{\Gamma_d}\frac{\Gamma_d^2}{\Gamma_d^2 + \delta^2}, \tag{5}$$

and the bath correlation time follows from $1/\tau_c = \Sigma R_{flip}$. Equation (5) is sharply peaked at δ = 0: pairs whose hyperfine energies coincide flip-flop freely, while pairs detuned by tens of megahertz are suppressed by $(\Gamma_d/\delta)^2 \approx 10^{-3}$. The bath dynamics are therefore controlled, to leading order, by the fraction of pairs that are hyperfine-degenerate.

This fraction is a pure counting problem. Two P1 centers are degenerate when $A_{eff}(i)\, m_I(i) = A_{eff}(j)\, m_I(j)$. Since the on-axis and off-axis values of $A_{eff}$ are incommensurate, this requires either that both nuclear spins are in the $m_I = 0$ state, or that the two centers share the same JT class and the same $m_I$. The probability that two randomly chosen P1 centers fall in the same JT class is $(1/4)^2 + (3/4)^2 = 5/8$. Hence

$$f_{\text{res}}(^{14}\text{N}) = \left(\frac{1}{3}\right)^2 + \frac{5}{8}\cdot\frac{2}{9} = \frac{1}{9} + \frac{5}{36} = \frac{1}{4}, \tag{6a}$$

$$f_{\text{res}}(^{15}\text{N}) = \frac{5}{8}\cdot\frac{1}{2} = \frac{5}{16}. \tag{6b}$$

The ratio is exactly 5/4. The $^{14}N$ case receives an extra contribution because the $m_I = 0$ state is degenerate with itself irrespective of the JT axes — a channel with no $^{15}N$ analogue — but this does not offset the larger probability, 1/2 versus 1/3, that two $^{15}N$ nuclei share the same $m_I$. A $^{15}N$ bath therefore has 25 % more resonant flip-flop channels than a $^{14}N$ bath, and a correspondingly faster bath. We stress that Eq. (6) contains no adjustable parameters: it follows from the nuclear spin multiplicity and the four-fold JT degeneracy alone.

For a single bath configuration the NV decays as $\exp[-(t/T_{2}^{*}{}_{\text{single}})^2]$ (Ramsey) and $\exp[-(t/T_{2,\text{single}})^3]$ (echo). Averaging over the spatial disorder of the bath requires the distributions of $\Delta_{single}$ and $\tau_{c,single}$, which Ref. [2] gives as

$$P(\Delta_{\text{single}}) = \frac{\Delta_{\text{ens}}}{\Delta_{\text{single}}^2}\sqrt{\frac{2}{\pi}}\exp\left(-\frac{\Delta_{\text{ens}}^2}{2\Delta_{\text{single}}^2}\right), \tag{7}$$

$$P(\tau_c) = \sqrt{\frac{\lambda}{2\pi\tau_c^3}}\exp\left[-\frac{\lambda(\tau_c-\tau_{\text{ens}})^2}{2\tau_c\tau_{\text{ens}}^2}\right], \tag{8}$$

the latter an inverse Gaussian (Wald) distribution. Integrating the single-NV decays over Eqs. (7) and (8) yields simple exponential Ramsey decay and stretched-exponential echo decay with $p \approx 3/2$, and identifies the ensemble times

$$T_2^* = \frac{1}{\Delta_{\text{ens}}}, \qquad T_2 = \left(\frac{2\tau_{\text{ens}}}{\Delta_{\text{ens}}^2}\right)^{1/3}. \tag{9}$$

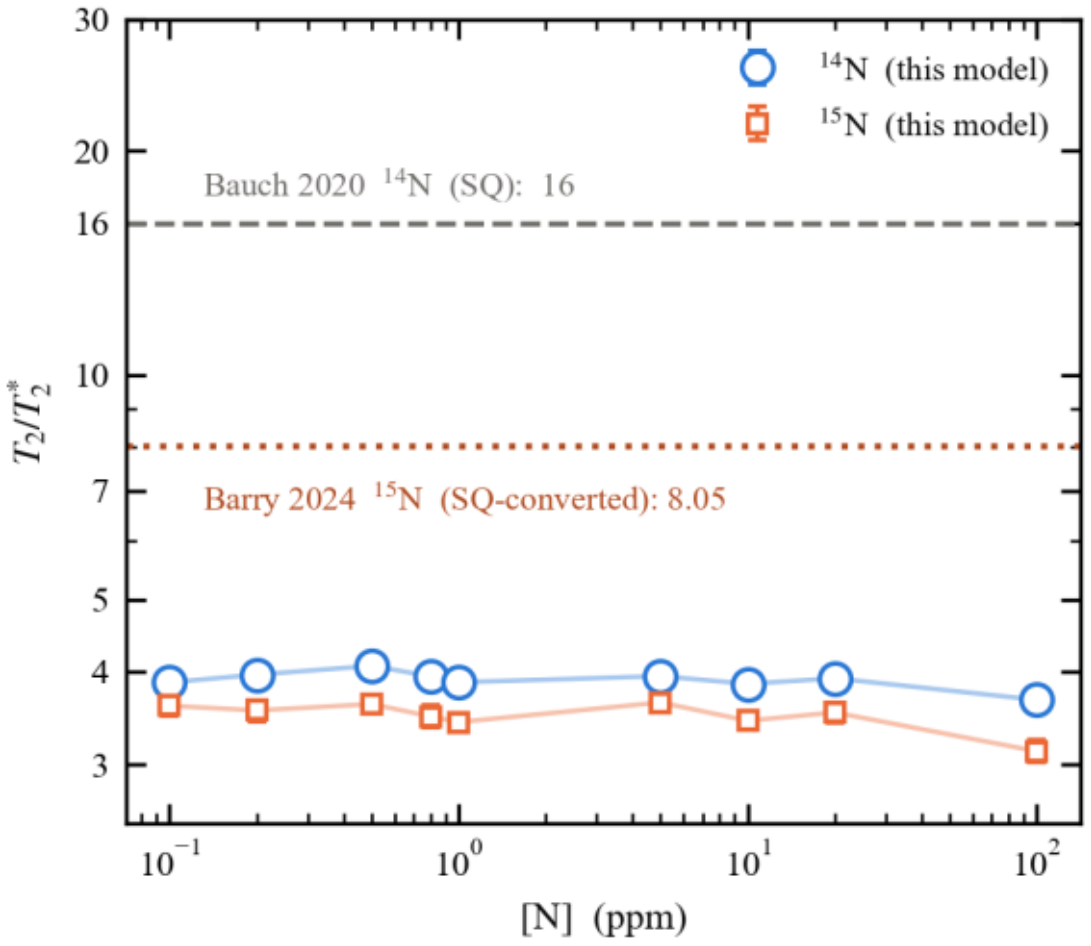

FIG. 2. Concentration-independent ratio $T_2/T_2^*$ for the two isotopes. The model gives a flat ratio, as required by the common 1/[N] scaling, with the $^{15}$N value lying systematically below the $^{14}$N value. Horizontal lines mark the experimental $^{14}$N value of Ref. [2] (dashed) and the $^{15}$N value of Ref. [9] converted to the same single-quantum convention (dotted). The model reproduces the ordering and the concentration-independence but not the full magnitude of the separation.

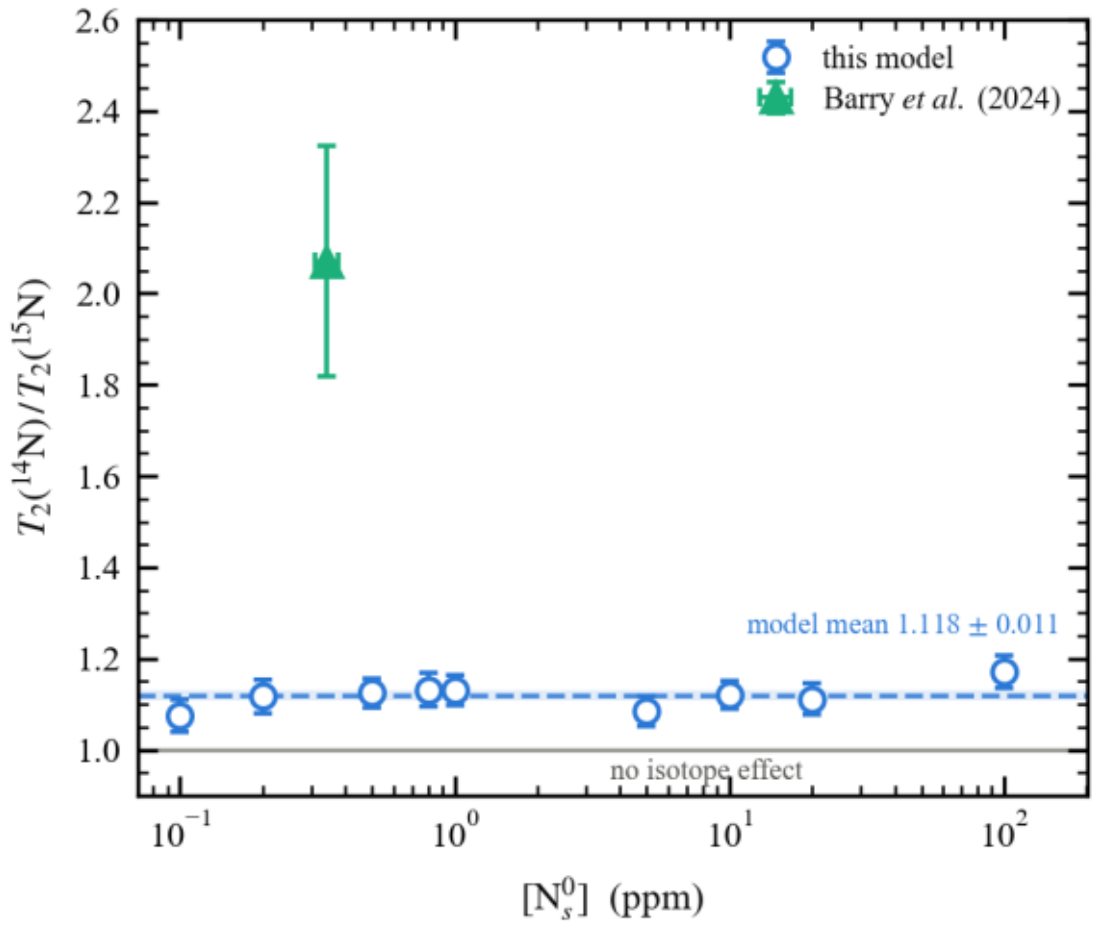

FIG. 3. Isotope ratio $T_2(^{14}\text{N})/T_2(^{15}\text{N})$ versus nitrogen concentration. Open circles are the semiclassical model of this work, with 1σ paired-bootstrap uncertainties; the dashed line and shaded band are their weighted mean, 1.118 ± 0.011. The filled triangle is the $^{15}$N magnetometer of Ref. [9]; the horizontal bar is the uncertainty in [N], which is itself inferred from the measured $T_2^*$ and the dephasing relation of Ref. [2]. The $^{14}$N coherence time entering the measured ratio is taken from the empirical relation 160 ± 12 μs·ppm of Ref. [2], whose 7.5 % uncertainty is included in the plotted error bar. The grey line at unity marks the absence of any isotope effect.

Equation (9) makes the asymmetry explicit: $T_2^*$ depends only on $\Delta_{\text{ens}}$, which is a property of the P1 positions and is blind to the nuclear isotope, whereas $T_2$ depends additionally on $\tau_{\text{ens}}$, which Eq. (6) shows is isotope-dependent. We extract $\Delta_{\text{ens}}$ and $\tau_{\text{ens}}$ by fitting Eqs. (7) and (8) to the simulated histograms and propagate the fit covariances to $T_2^*$ and $T_2$. Implementation details, including the unit convention, the treatment of $\Gamma_d$, and the numerical acceleration that makes the 0.1 ppm point tractable, are given in the Supplementary Material.

Figure 1 shows the simulated $T_2^*$ and $T_2$ over three decades of nitrogen concentration. Fitting $T = A/[\text{N}]^p$ with weights from the propagated errors gives $p = 1.005 \pm 0.003$ for $T_2^*$ and $p = 1.013 \pm 0.003$ ($^{14}$N), $p = 1.020 \pm 0.003$ ($^{15}$N) for $T_2$ — inverse-linear scaling within a few parts per thousand, in agreement with the experimental observation of Ref. [2]. Constraining $p = 1$ gives the slopes collected in Table I.

TABLE I. Coherence-time slopes from the constrained fit $T = A/[\text{N}]$, compared with the measured values of Ref. [2]. Units of A are μs·ppm; Ratio is model/experiment.

| | Isotope | A (model) | A (expt.) [2] | Ratio |
|---|---|---|---|---|
| $T_2^*$ | $^{14}$N and $^{15}$N | 11.21 ± 0.08 | 9.6 ± 0.9 | 1.17 |
| $T_2$ | $^{14}$N | 44.06 ± 0.31 | 160 ± 12 | 0.28 |
| $T_2$ | $^{15}$N | 39.28 ± 0.27 | — | — |

The simulated $T_2^*$ slope, 11.21 ± 0.08 μs·ppm, agrees with the measured 9.6 ± 0.9 μs·ppm to within 17 %, i.e. within roughly two standard deviations of the experimental uncertainty, and involves no fitting parameters. The simulated $T_2$ slope is a factor 3.6 below the measured value, an underestimate already noted and discussed in Ref. [2], whose authors attribute it to physics beyond the mean-field stochastic treatment.

The isotope effect appears entirely in the bath dynamics. Across all nine concentrations the fitted correlation time is longer for $^{14}$N by a factor 1.40 (range 1.24–1.61), consistent with the 5/4 resonant-channel ratio of Eq. (6) once the skew of the inverse-Gaussian distribution is taken into account (Supplementary Material, Sec. S1 G). Because $T_2 \propto \tau_{\text{ens}}^{1/3}$, this compresses to a $T_2$ ratio of 1.122, while $T_2^*$ is unchanged.

Figure 2 collects the result. The model gives $T_2/T_2^* = 3.93$ for $^{14}$N and 3.50 for $^{15}$N, a reduction by a factor 1.12, flat in concentration as required. Experimentally the corresponding numbers are 16 [2] and 8.05 [9], a reduction by 1.99.

The cleanest way to expose the isotope effect is to plot the ratio $T_2(^{14}\text{N})/T_2(^{15}\text{N})$ directly, since every quantity that is common to the two isotopes — the P1 positions, $\Delta_{ens}$, and hence $T_2^*$ — cancels exactly. Because our two isotope runs use identical random seeds, they share the same P1 configurations trial by trial, so the ratio can be evaluated as a paired quantity and its uncertainty obtained by paired bootstrap resampling. This is roughly an order of magnitude tighter than propagating the two fit errors as if they were independent. Figure 3 shows the result.

Three statements follow from Fig. 3, and we state them in order of how firmly the calculation supports them.

First, the simulated ratio is independent of nitrogen concentration. A weighted fit of a power law to the nine simulated points gives an exponent of $+0.0055 \pm 0.0048$, consistent with zero at $1.1\sigma$, and a constant fit describes the points with $\chi^2/\text{dof} = 0.65$. This is not an accident of the numerics but a structural property of the model: Eq. (6) counts resonant channels from the nuclear multiplicity and the Jahn-Teller degeneracy alone, neither of which depends on [N], while every geometric factor in Eqs. (2), (3) and (5) is common to the two isotopes. The ratio is therefore a genuine material constant of the nitrogen isotope, not a sample-dependent number, which is what makes it a meaningful target for experiment.

Second, the ratio is definitively greater than unity. The weighted mean over the nine concentrations is $1.118 \pm 0.011$, which exceeds 1 by $11\sigma$, and all nine individual points lie above unity. A $^{15}N$ bath is unambiguously the faster bath, and $^{15}N$ enrichment therefore carries an intrinsic penalty in $T_2$ at fixed nitrogen concentration — a conclusion of direct practical relevance, since $^{15}N$ doping is chosen in ensemble magnetometers for the simplification of the microwave spectrum [9], not for its coherence properties.

Third, and most importantly for what comes next, the measured ratio lies above the calculation. Combining the $^{15}N$ magnetometer of Ref. [9] with the empirical $^{14}N$ relation of Ref. [2] gives $2.07 \pm 0.25$ at [N] = 0.34 ppm, against the simulated $1.118 \pm 0.011$. The offset is +0.95, or $3.8\sigma$ for the quasi-static basis conversion, on the combined uncertainty, and the measured value is a factor 1.85 larger than the model. The uncertainty is dominated entirely by the experimental side — 9.4 % from the dephasing slope of Ref. [2] that sets [N], 7.5 % from its $^{14}N$ coherence slope, and 2.1 % from the measured $T_2$ of Ref. [9] — while the simulated ratio is determined to 1 %. The discrepancy is therefore significant for the adopted conversion but not yet sharply bounded, and closing it is a matter of both better calculation and better measurement.

Three findings should be separated. First, the model reproduces $T_2^*$ essentially quantitatively and predicts, for a structural reason rather than by fitting, that $T_2^*$ is insensitive to the nitrogen isotope: Eq. (3) contains only P1 positions. An isotope-dependent $T_2^*$ at fixed [N] would indicate physics outside this framework.

Second, the sign and the concentration-independence of the isotope effect on $T_2$ are reproduced by a parameter-free counting argument, 5/16 versus 1/4 resonant channels. It is instructive to contrast this with the host-lattice case. For a defect spin in a dilute nuclear bath, the scaling of Ref. [4] gives a coherence time that decreases with the nuclear spin of the bath species, roughly as $I^{-1.1}$. In the P1 bath the dependence runs the other way: the nitrogen nucleus is not itself a noise source on the timescale of the echo [5] but only detunes the electron flip-flops, and a larger I spreads the transition frequencies over more values and slows the bath. The same isotope substitution thus shortens coherence in one kind of bath and lengthens it in the other, depending on whether the nucleus belongs to the bath or is merely attached to it.

Third, the magnitude falls short: the model reduction of 1.12 accounts for roughly 17 % of the observed one on a logarithmic measure. (i) The rate equation (5) treats each P1 pair independently with a mean dephasing rate $\Gamma_d$, which Ref. [2] already identifies as the origin of its threefold underestimate of $T_2$; the CCE calculations of Ref. [5] show that hyperfine suppression of flip-flops is considerably stronger than a rate-equation picture suggests. (ii) The two experimental numbers come from different laboratories and growth runs, and the nitrogen concentration of Ref. [9] is inferred from its own $T_2^*$. (iii) The DQ-to-SQ conversion is the one model-dependent step: the measured ratio would be reproduced by the model for a conversion factor of about 2.9 instead of 1.59, and the significance of the discrepancy is conditional on this choice (Supplementary Material, Sec. S4).

We therefore regard the counting argument as a lower bound on the isotope contribution. Because the simulated plateau in Fig. 3 is flat by construction, the missing factor of 1.85 must be a concentration-independent enhancement of the contrast, and CCE is the natural candidate. In a many-body treatment a detuned pair is not simply suppressed by $(\Gamma_d/\delta)^2$ but still contributes through its correlated evolution. Since $^{14}N$ has five distinct hyperfine energies — $m_I \in \{-1, 0, +1\}$ combined with the on-axis and off-axis $A_{eff}$ values — against four for $^{15}N$, a treatment sensitive to near-degenerate channels should produce a larger contrast than channel counting alone; whether it produces enough is the question a CCE calculation on this system is designed to answer.

The most direct experimental test is a pair of diamonds grown in the same reactor and differing only in nitrogen isotope, measured in the same apparatus and basis. This would remove the 9.4 % and 7.5 % uncertainties that the empirical $^{14}N$ relations of Ref. [2] contribute to the measured point, the inter-laboratory systematic and the basis ambiguity at once. Our prediction is unambiguous: identical $T_2^*$ at matched [N], and $T_2$ shorter in the $^{15}N$ sample by a factor of at least 1.12.

We have shown that the nitrogen nuclear isotope controls the spin coherence of the $NV^-$ ensemble — $T_2$ but not $T_2^*$ — through a purely combinatorial mechanism: the fraction of P1 pairs that are hyperfine-degenerate, and hence free to flip-flop, is 1/4 for $^{14}N$ and 5/16 for $^{15}N$. Within the semiclassical model of Ref. [2], augmented with the Jahn-Teller-resolved hyperfine tensor of Ref. [5], this shortens the bath correlation time by 1.40, shortens $T_2$ by 1.122, and leaves $T_2^*$ exactly unchanged, lowering $T_2/T_2^*$ from 3.93 to 3.50. Expressed as the isotope ratio, the model predicts $T_2(^{14}N)/T_2(^{15}N) = 1.118 \pm 0.011$, a genuine constant: independent of nitrogen concentration over three decades (power-law exponent +0.0055 ± 0.0048) and greater than unity by 11σ. Measurement places the contrast higher still, at 2.07 ± 0.25 when the $^{15}$N magnetometer of Ref. [9] is referred to the empirical $^{14}$N relation of Ref. [2], a 3.8σ offset, for the quasi-static basis conversion, from a plateau that the model fixes to 1 %. This reproduces the direction and the concentration-independence of the isotope effect, though not its full magnitude. Because the channel count does not depend on nitrogen concentration, the same prediction applies to the parts-per-billion P1 baths that now appear to limit the coherence of single NV centers [8]. A cluster-correlation-expansion treatment of the same system, in which near-degenerate as well as exactly degenerate flip-flop channels contribute, is the natural next step and the one we expect to close the gap.

*Acknowledgments*—This work was supported by JSPS KAKENHI Grant No. JP25K17636.

*Data availability*—The data that support the findings of this article are not publicly available. The data, including the numerical values plotted in all figures and tables, and the simulation code are available from the author upon reasonable request.

---

# Supplementary Material for Semiclassical spin-bath calculation of the nitrogen-isotope effect on $NV^-$ ensemble coherence in diamond

Chikara Shinei[1,2,*]

[1]Department of Applied Physics, Institute of Pure and Applied Sciences, University of Tsukuba, Tsukuba, Ibaraki 305-8573, Japan
[2]Japanese-French Laboratory for Semiconductor Physics and Technology J-FAST, CNRS, University Grenoble Alpes, Grenoble INP, University of Tsukuba, Tsukuba, Japan
[*]shinei.chikara.fb@u.tsukuba.ac.jp

## S1. IMPLEMENTATION OF THE SEMICLASSICAL MODEL

This section documents the choices required to turn the equations of Bauch et al. [1] into a numerically stable and convergent simulation. Several of them are not stated explicitly in the original paper but change the results by large factors, so we record them in full. The simulation code accompanies this manuscript.

### *S1 A. Unit convention: angular versus ordinary frequency*

The quantities $\Delta$, $\Omega$, $\Gamma_d$ and $\delta$ in Ref. [1] are angular frequencies (rad/μs), not ordinary frequencies. This is easy to get wrong and costs a factor of $2\pi$ in every derived time. The convention can be fixed unambiguously from Ref. [1] itself: Fig. 4(a) of that work shows the $\Delta_{single}$ distribution peaking near 7 $\mu s^{-1}$ at [N] = 100 ppm, and the mode of Eq. (7) is $\Delta_{ens}/\sqrt{2}$, giving $\Delta_{ens} \approx 10$ rad/μs. This is exactly $2\pi \times 16$ kHz/ppm × 100 ppm, i.e. the measured dephasing rate A(NV-N)·[N]. Accordingly the dipolar coefficient is used as $(\mu_0/4\pi)\hbar\gamma_e^2 = 2\pi \times 52.04 = 327.0$ rad·MHz·nm³, and the hyperfine constants are multiplied by $2\pi$ before entering Eqs. (5) and (3) of the main text. Using ordinary frequencies throughout instead inflates $T_2^*$ by $2\pi$ and corrupts the flip-flop rate by the same factor.

### *S1 B. Secular hyperfine energy for off-axis Jahn-Teller orientations*

A P1 center whose JT axis is not parallel to the applied field does not have secular hyperfine splitting $A_\perp$. The correct value is the projection of the axially symmetric tensor onto the field direction, Eq. (4) of the main text. For $^{14}N$ this gives 85.57 MHz rather than 81.31 MHz for the three off-axis orientations. The resulting on-axis/off-axis difference of 28.5 MHz agrees with the ≈ 29 MHz level shift reported in the CCE analysis of Ref. [2], which we take as an independent validation.

### *S1 C. The bath linewidth $\Gamma_d$*

Reference [1] prescribes $\Gamma_d \approx \sqrt{N_b} \cdot \bar{C}_\parallel$, where Nb is the number of bath spins and $\bar{C}_\parallel$ the average dipolar interaction. Evaluated literally as an average over all pairs inside a simulation sphere of radius R, this expression is not convergent: $\sqrt{N_b}$ grows as $R^{3/2}$ while $\bar{C}_\parallel$ falls as $R^{-3}$, so $\Gamma_d \propto R^{-3/2}$. We measured 2.9, 1.8, 1.5 and 1.1 MHz at R = 12, 16, 20 and 24 nm for [N] = 100 ppm, i.e. no convergence at all.

Since $\Gamma_d$ is defined in Ref. [1] as "the intrinsic linewidth of the dipolar spin bath", we instead evaluate it directly as that linewidth. For each configuration we draw a random Ising bath state $\sigma = \pm 1$ and compute the local field on every P1 center, then take

$$\Gamma_d = \underset{i}{\text{median}} \left| \sum_j \frac{C_\parallel^{ij} \sigma_j}{2} \right|. \tag{S1}$$

A dilute dipolar spin system has a Lorentzian line, for which the median absolute local field is the half-width at half-maximum; the median is also robust against the divergent second moment that makes $\sqrt{\Sigma C_\parallel^2}$ unusable. Equation (S1) is convergent in R (1.54, 1.60, 1.67, 1.72 MHz at R = 12, 16, 20, 26 nm for [N] = 100 ppm) and scales linearly with concentration (1.6, 0.47, 0.16 MHz at 100, 30, 10 ppm), as a dipolar linewidth must.

### *S1 D. Convergent evaluation of the correlation time*

Equation (5) of the main text summed literally over all P1 pairs, $1/\tau_c = \Sigma R_{flip}$, diverges: the number of pairs grows as $N^2$ while the individual rates fall only as $r^{-6}$, so the sum grows without bound with the simulation radius. We measured $\tau_c = 0.29$, 0.008 and 0.002 ns at R = 12, 20 and 24 nm for [N] = 100 ppm — three to five orders of magnitude below the ≈ 1 μs of Ref. [1], and yielding the unphysical result $T_2 < T_2^*$.

Appendix D of Ref. [1] states that "we ignore spin bath pairs that interact weakly with the NV, leading to motional narrowing", but does not specify the cutoff. Rather than introduce an arbitrary cutoff radius, we weight each pair by how much its flip-flop actually changes the field at the NV. A flip-flop of the pair (i, j) changes the NV field by $C_\parallel(NV,i) - C_\parallel(NV,j)$, so

$$\frac{1}{\tau_c} = \sum_{(ij)} R_{flip}^{ij} \frac{\left[C_\parallel^{NV,i} - C_\parallel^{NV,j}\right]^2}{4\Delta_{single}^2}. \tag{S2}$$

The weight falls as $\rho^{-6}$ with the distance $\rho$ of the pair from the NV while the number of such pairs grows only as $\rho^2$, so Eq. (S2) converges. It is parameter-free and reduces to the literal sum for a bath in which all spins couple equally to the NV.

### *S1 E. Saturation of the flip-flop rate*

Equation (5) of the main text is a perturbative result valid for $\Omega \ll \Gamma_d$. For the closest P1 pairs $\Omega$ reaches $10^4$ rad/μs and the expression diverges as $\Omega^2/\Gamma_d$, producing $\tau_c$ values of order 10 ps — faster than the bath linewidth permits. The exact Bloch-equation solution for a two-level system with coupling $\Omega$, detuning $\delta$ and dephasing $\Gamma_d$ contains a saturation term:

$$R_{\mathrm{flip}} = \frac{\Omega^2\Gamma_{\mathrm{d}}}{\Gamma_{\mathrm{d}}^2 + \delta^2 + 2\Omega^2}. \tag{S3}$$

Equation (S3) reduces to Eq. (5) when $\Omega \ll \Gamma_d$ and caps the rate at $\Gamma_d/2$ otherwise. Including it removes the unphysical tail (the minimum $\tau_c$ rises from 0.1 ns to 0.1 μs at [N] = 100 ppm) and, importantly, is what makes the simulated $\tau_c$ distribution take the inverse-Gaussian form assumed by Eq. (8); without it the distribution spans seven decades and Eq. (8) does not describe it. We regard this as strong internal evidence that the saturation term is required for consistency with Ref. [1]'s own ansatz.

### *S1 F. Estimators for the ensemble parameters*

Both distributions have heavy $1/x^2$ tails, so the sample mean of $\Delta_{\mathrm{single}}$ and the sample median of $\tau_c$ are poor estimators. Two exact results simplify the analysis. First, substituting $u = 1/\Delta$ into Eq. (7) turns it into a half-normal distribution of scale $1/\Delta_{\mathrm{ens}}$, so the maximum-likelihood estimator is available in closed form:

$$\Delta_{\mathrm{ens}} = \frac{1}{\sqrt{\langle \Delta_{\mathrm{single}}^{-2} \rangle}}. \tag{S4}$$

The same substitution gives median($\Delta_{\mathrm{single}}$) = 1.4826 × $\Delta_{\mathrm{ens}}$, so using the raw median as $\Delta_{\mathrm{ens}}$ overestimates $T_2^*$ by 48 %. Second, the mean parameter of the inverse Gaussian, Eq. (8), has the sample mean as its maximum-likelihood estimator.

For the values quoted in the main text we nevertheless follow the procedure of Ref. [1] and fit Eqs. (7) and (8) to the simulated histograms, because that is what provides the fit covariance matrix from which the quoted uncertainties are propagated. One modification is necessary: fitting in density space lets the first one or two bins dominate the residual, and the inverse-Gaussian fit then degenerates towards $\lambda \to 0$, returning meaningless values (we observed $\tau_{\mathrm{ens}}$ = 9089 μs where the true value was 2 μs). Fitting log(density) instead weights all bins equally and is stable. Table S1 reports both the fitted and the maximum-likelihood values; their agreement is a check that the simulated distributions really do follow Eqs. (7) and (8).

### *S1 G. Relation between the channel count and the simulated correlation time*

Equation (6) of the main text predicts a resonant-channel ratio of exactly 5/4. The quantity that actually enters the rate is the Lorentzian weight $\Gamma_d^2/(\Gamma_d^2 + \delta^2)$ averaged over pairs. Evaluating this directly in the simulation gives 0.2499 ($^{14}$N) and 0.3115 ($^{15}$N) at [N] = 1 ppm, and 0.2517 and 0.3151 at 100 ppm, i.e. ratios of 1.247 and 1.252 — the analytic 5/4 to three digits, and independent of concentration. Correspondingly the ratio of $\langle 1/\tau_c \rangle$ is 1.23 (1 ppm) and 1.31 (100 ppm).

The fitted $\tau_{\mathrm{ens}}$ ratio quoted in the main text, 1.40, is somewhat larger than 5/4 because Eq. (8) is strongly right-skewed: the mean of the distribution is not the reciprocal of the mean rate. The chain 5/4 → $\langle 1/\tau_c \rangle$ → $\tau_{\mathrm{ens}}$ → $T_2 \propto \tau_{\mathrm{ens}}^{1/3}$ therefore ends at a $T_2$ ratio of 1.122.

### *S1 H. Paired bootstrap for the isotope ratio*

The two isotope runs at each concentration use identical random seeds, so trial k of the $^{14}$N run and trial k of the $^{15}$N run use the same P1 positions and the same Jahn-Teller axes; only the nuclear spin multiplicity differs. This makes $\Delta_{\mathrm{ens}}$ bitwise identical between the isotopes — which is why the $T_2^*$ points in Fig. 1(a) coincide exactly — and it makes the isotope ratio a paired quantity:

$$\frac{T_2(^{14}\mathrm{N})}{T_2(^{15}\mathrm{N})} = \left[\frac{\tau_{\mathrm{ens}}(^{14}\mathrm{N})}{\tau_{\mathrm{ens}}(^{15}\mathrm{N})}\right]^{1/3}. \tag{S5}$$

The spatial disorder of the bath, which dominates the uncertainty on either coherence time separately, therefore cancels in the ratio. We exploit this by resampling the two correlation-time samples with a common index vector: for each of 400 bootstrap replicates we draw 2000 trial indices with replacement, apply the same indices to both isotopes, refit Eq. (8) to each resampled histogram exactly as in the main analysis, and form the ratio. The quoted uncertainties in Fig. 3 are the standard deviations of those replicates. They are about an order of magnitude smaller than what independent propagation of the two fit errors would give, and they are the statistically correct ones, because the two runs are not independent by construction.

The weighted mean of the nine simulated ratios is 1.1184 ± 0.0112 with $\chi^2$/dof = 5.17/8 = 0.65, i.e. the points are consistent with a single constant. A weighted power-law fit gives an exponent of +0.0055 ± 0.0048, consistent with zero at 1.1σ. The ratio exceeds unity by 11σ.

*S1 I. Sampling of P1 positions and computational cost*

The straightforward implementation — build the full diamond lattice inside the simulation sphere, then select each site as a P1 with probability [N] — becomes impossible at low concentration. At [N] = 0.1 ppm the mean P1 separation is 23.9 nm, so a sphere large enough for convergence has R ≈ 167 nm and contains $2.7 \times 10^8$ carbon sites; the coordinate array alone exceeds 6 GB.

We avoid this entirely by sampling the P1 positions directly. The number of P1 centers in the sphere is drawn from a Poisson distribution with mean $n_C \cdot V \cdot [N]$ (the Poisson limit of the binomial is exact to many digits here, since the site count is enormous and the occupation probability minute). Each position is then obtained by drawing a unit-cell index and a basis index uniformly from the bounding cube and rejecting points outside the sphere, which is equivalent to choosing N sites uniformly at random from those inside it. The acceptance rate is the sphere-to-cube volume ratio, about 52 %.

The cost of this scheme depends only on the number of P1 centers, not on the lattice size. We therefore set the simulation radius to a fixed multiple of the mean P1 separation, $R = 7 \times (3/4\pi n)^{1/3}$, so that every concentration contains $\approx 7^3 = 343$ P1 centers. Two benefits follow. First, the runtime is identical at every concentration: 0.1 ppm (R = 167 nm) costs the same 15 ms per configuration as 100 ppm (R = 16.7 nm), against which the lattice-based scheme could not reach 0.1 ppm at all. Second, the finite-size systematic is the same at every concentration and therefore cancels from the scaling exponent p and from the $T_2/T_2^*$ ratio, which are the quantities this work is about.

All pairwise quantities are evaluated as vectorized N × N array operations rather than Python loops, which is a further factor of order $10^2$ in speed. The complete sweep of nine concentrations × two isotopes × 2000 configurations runs in about nine minutes on a single core.

*S1 J. Other numerical details*

The detuning δ includes, as stated in Ref. [1], the P1 nuclear spin state, the P1 JT axis, and the Ising fields of neighbouring P1 centers (excluding the mutual term of the pair under consideration). Nuclear spin states and JT axes are drawn uniformly and are static within a configuration. The bath Ising state used for $\Gamma_d$ and for the Overhauser field is the same draw. $\sin^2\theta$ is clipped to [0, 1] to guard against round-off. All results use 2000 configurations per point; the reported uncertainties are dominated by the fit, not by the configuration count.

## S2. DISTRIBUTIONS AND FITS AT EACH CONCENTRATION

Figures S1 and S2 show the simulated distributions of $\Delta_{single}$ and $\tau_{c,single}$ at each nitrogen concentration, together with the fitted Eqs. (7) and (8); both figures are collected at the end of this document. Histograms are plotted up to the smaller of five times the fitted scale parameter and the 90th percentile of the sample; the fits themselves use the same range but all 2000 configurations contribute.

The insets of Fig. S2 isolate the quantity that moves. The mode of the inverse Gaussian of Eq. (8) is available in closed form,

$$\mathrm{mode}(\tau_c) = \tau_{\mathrm{ens}} \left[ \sqrt{1 + \frac{9\tau_{\mathrm{ens}}^2}{4\lambda^2}} - \frac{3\tau_{\mathrm{ens}}}{2\lambda} \right], \tag{S6}$$

and because the two isotopes share the same P1 configurations the two modes can be compared directly. Averaged over the nine concentrations the ratio of the modes is 1.35, with a median of 1.34 and a spread of 1.25–1.62; excluding the lowest concentration, where the fit is least constrained, the values lie between 1.25 and 1.38. This is close to the 5/4 = 1.25 predicted by the resonant-channel count of Eq. (6), and is a more direct read-out of that count than the fitted mean $\tau_{ens}$, which is inflated by the long tail of the distribution.

## S3. TABULATED FIT VALUES

Table S1 lists every fitted parameter with its 1σ uncertainty. Uncertainties on $T_2^*$ and $T_2$ are obtained from the fit covariance matrix by the standard propagation

$$\sigma(T_2^*) = \frac{\sigma(\Delta_{\mathrm{ens}})}{\Delta_{\mathrm{ens}}^2}, \qquad \sigma(T_2) = \frac{T_2}{3}\sqrt{\left(\frac{\sigma(\tau_{\mathrm{ens}})}{\tau_{\mathrm{ens}}}\right)^2 + 4\left(\frac{\sigma(\Delta_{\mathrm{ens}})}{\Delta_{\mathrm{ens}}}\right)^2}, \tag{S7}$$

treating the Δ and τ fits as independent. We emphasise that these are fit uncertainties only, of order 2–3 %. They do not include the systematic uncertainty of the model itself, which is far larger: the choices documented in Secs. S1 C–S1 E move $\tau_{ens}$ by factors of two to three, and the model underestimates the absolute $T_2$ by 3.6× relative to experiment. The error bars in Figs. 1 and 2 should be read as statistical precision, not accuracy.

TABLE S1. Fitted ensemble parameters and derived coherence times. Each row is 2000 Monte Carlo configurations.

| [N] (ppm) | Iso. | $\Delta_{ens}$ (rad/μs) | $\sigma(\Delta_{ens})$ | $\tau_{ens}$ (μs) | $\sigma(\tau_{ens})$ | $T_2^*$ (μs) | $\sigma(T_2^*)$ | $T_2$ (μs) | $\sigma(T_2)$ | $T_2/T_2^*$ |
|---|---|---|---|---|---|---|---|---|---|---|
| 0.1 | 14N | 0.009024 | 0.00018 | 3209 | 1.6e+02 | 110.8 | 2.2 | 428.7 | 9.2 | 3.87 |
| 0.1 | 15N | 0.009024 | 0.00018 | 2584 | 1.6e+02 | 110.8 | 2.2 | 398.9 | 9.6 | 3.60 |
| 0.2 | 14N | 0.0172 | 0.00044 | 1809 | 73 | 58.15 | 1.5 | 230.4 | 5 | 3.96 |

| [N] (ppm) | Iso. | $\Delta_{ens}$ (rad/μs) | $\sigma(\Delta_{ens})$ | $\tau_{ens}$ (μs) | $\sigma(\tau_{ens})$ | $T_2^*$ (μs) | $\sigma(T_2^*)$ | $T_2$ (μs) | $\sigma(T_2)$ | $T_2/T_2^*$ |
|---|---|---|---|---|---|---|---|---|---|---|
| 0.2 | 15N | 0.0172 | 0.00044 | 1298 | 43 | 58.15 | 1.5 | 206.3 | 4.2 | 3.55 |
| 0.5 | 14N | 0.04633 | 0.00082 | 728.1 | 41 | 21.58 | 0.38 | 87.86 | 2 | 4.07 |
| 0.5 | 15N | 0.04633 | 0.00082 | 511.1 | 23 | 21.58 | 0.38 | 78.08 | 1.5 | 3.62 |
| 0.8 | 14N | 0.06914 | 0.0017 | 443.2 | 22 | 14.46 | 0.36 | 57.02 | 1.4 | 3.94 |
| 0.8 | 15N | 0.06914 | 0.0017 | 305.5 | 14 | 14.46 | 0.36 | 50.37 | 1.1 | 3.48 |
| 1 | 14N | 0.08092 | 0.0016 | 358.2 | 13 | 12.36 | 0.25 | 47.83 | 0.88 | 3.87 |
| 1 | 15N | 0.08092 | 0.0016 | 247.6 | 9.9 | 12.36 | 0.25 | 42.29 | 0.8 | 3.42 |
| 5 | 14N | 0.46 | 0.0086 | 66.68 | 3.1 | 2.174 | 0.041 | 8.574 | 0.17 | 3.94 |
| 5 | 15N | 0.46 | 0.0086 | 52.24 | 2.2 | 2.174 | 0.041 | 7.904 | 0.15 | 3.64 |
| 10 | 14N | 0.9081 | 0.018 | 31.45 | 1.7 | 1.101 | 0.021 | 4.241 | 0.095 | 3.85 |
| 10 | 15N | 0.9081 | 0.018 | 22.35 | 0.94 | 1.101 | 0.021 | 3.785 | 0.072 | 3.44 |
| 20 | 14N | 1.817 | 0.04 | 16.55 | 0.73 | 0.5503 | 0.012 | 2.156 | 0.045 | 3.92 |
| 20 | 15N | 1.817 | 0.04 | 12.06 | 0.51 | 0.5503 | 0.012 | 1.94 | 0.04 | 3.53 |
| 100 | 14N | 9.026 | 0.24 | 2.728 | 0.14 | 0.1108 | 0.0029 | 0.4061 | 0.0098 | 3.67 |
| 100 | 15N | 9.026 | 0.24 | 1.695 | 0.065 | 0.1108 | 0.0029 | 0.3465 | 0.0075 | 3.13 |

## S4. CONVERSION FROM THE DOUBLE-QUANTUM TO THE SINGLE-QUANTUM CONVENTION

Reference [1] reports $T_2$ from single-quantum (SQ) spin echo, and reports $T_2^*$ from double-quantum (DQ) Ramsey corrected to the SQ convention by the factor of two that its Appendix B prescribes. The $^{15}N$ data of Ref. [3] are measured in the DQ basis [4] and must be converted before they can be compared. The two corrections are not the same number. A DQ measurement doubles the accumulated magnetic phase, $\Delta \to 2\Delta$. For a Ramsey measurement $T_2^* = 1/\Delta$ and the correction is exactly a factor of two. For a Hahn echo the exponent depends on the noise regime: in the quasi-static limit $t \ll \tau_c$ the coherence integral gives $\chi \propto \Delta^2 t^3$ and hence $T_2 \propto \Delta^{-2/3}$, a factor $2^{2/3} = 1.59$, whereas in the opposite, motionally narrowed limit $\chi \propto \Delta^2 \tau_c\, t$ would give $\Delta^{-2}$ and a factor 4.

The sample of Ref. [3] lies on the quasi-static side: at 0.34 ppm the bath correlation time is of order $10^3$ μs (Table S1), several times longer than the measured $T_2$. We therefore adopt the quasi-static factor 1.59 throughout, which gives $T_2^{*,SQ}$ = 28.0 μs, $T_{2,SQ}$ = 225.4 μs and $T_2/T_2^* = 8.05 \pm 0.18$ for the $^{15}N$ sample, a factor 1.99 below the $^{14}N$ value of 16.

This is the one model-dependent step in the comparison, and the measured isotope ratio is sensitive to it. With the quasi-static factor the measured ratio is 2.07; with the motionally narrowed factor of 4 it would be 0.82, below both the model and unity; the model value would be reproduced by an intermediate factor of about 2.9. The stretching exponent $p = 1.14$ fitted in Ref. [3], closer to the motionally narrowed value $p = 1$ than to the quasi-static ensemble value $p \approx 3/2$, is a reminder that the conversion is not settled; the significance of the discrepancy quoted in the main text is conditional on it. A measurement of both isotopes in the same basis would remove this step altogether.

## S5. DATA AVAILABILITY

The numerical data behind every figure and table are not included with this article: the simulated coherence times and power-law fits (Figs. 1 and 2), the isotope ratios and paired-bootstrap statistics (Fig. 3), the histograms, fitted distributions and modes (Figs. S1 and S2), the fitted parameters (Table S1), the full simulation output including the maximum-likelihood estimates used as a cross-check, and the per-configuration samples. They are available from the author upon reasonable request, as is the simulation code.

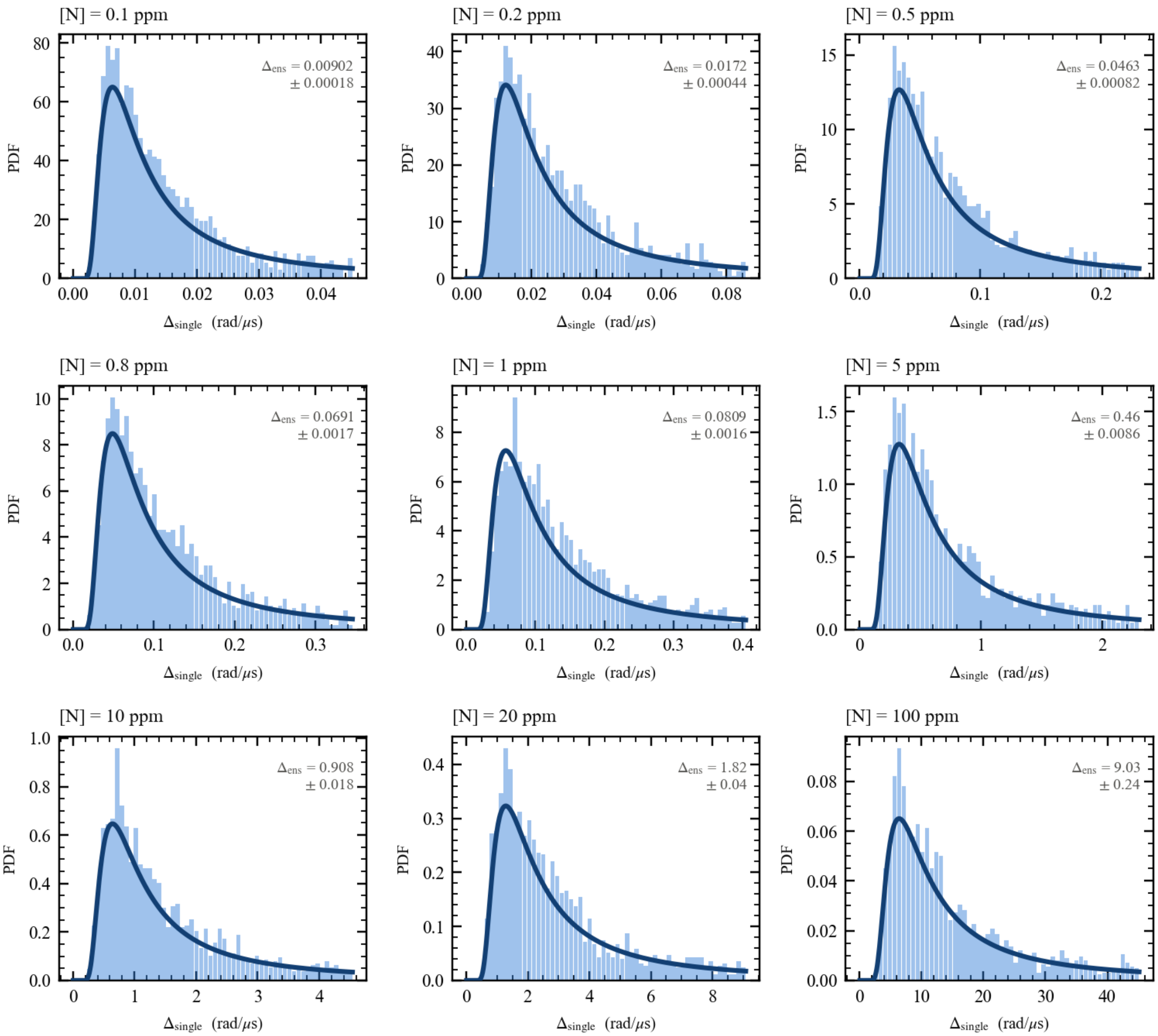

FIG. S1. Distribution of the NV–bath coupling $\Delta_{single}$ at each nitrogen concentration, with fits to Eq. (7). Fitted $\Delta_{ens}$ values and their 1σ uncertainties are annotated. These distributions are identical for $^{14}N$ and $^{15}N$ — Eq. (3) involves only the P1 positions — which is the origin of the isotope-independence of $T_2^*$.

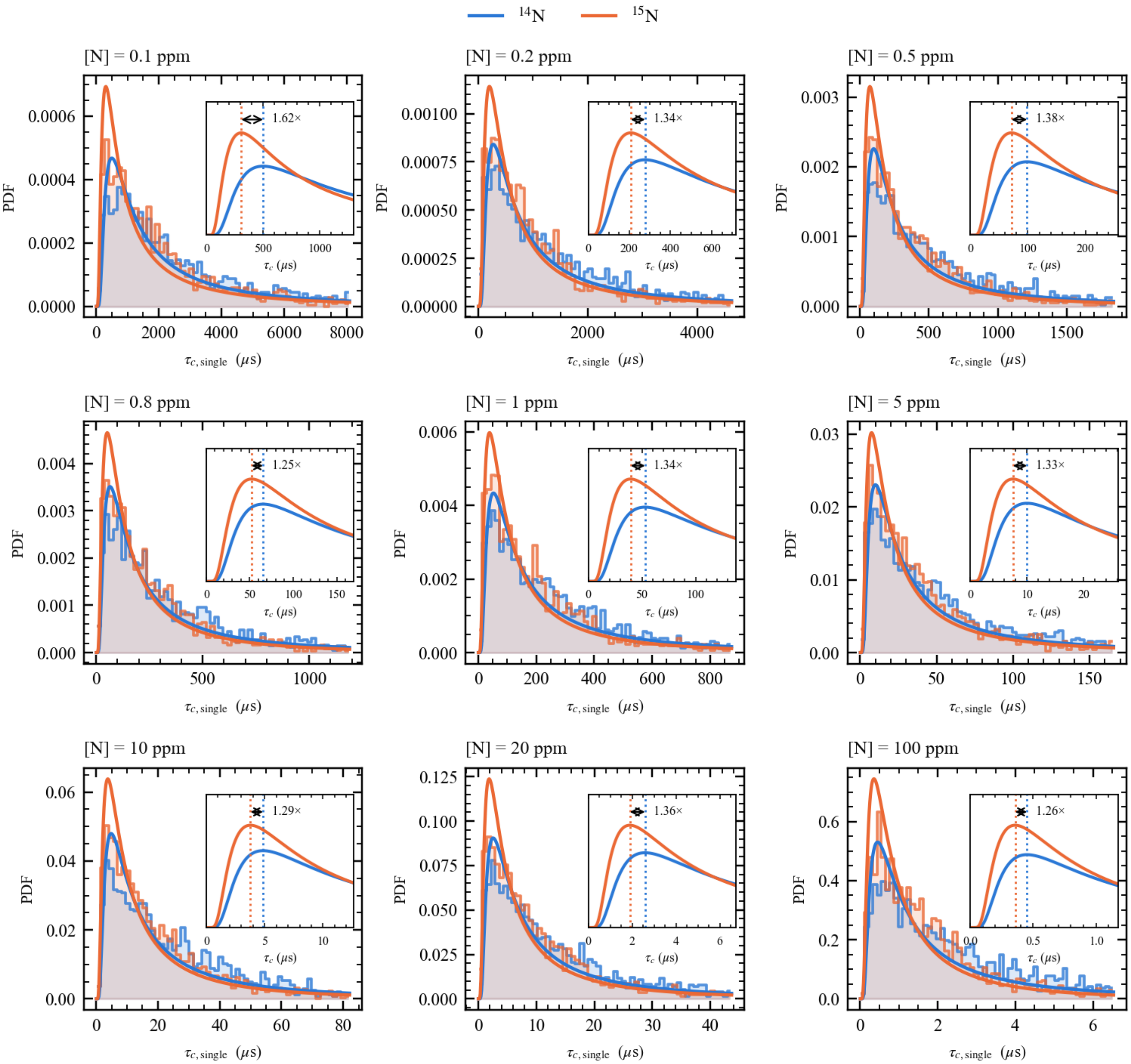

FIG. S2. Distribution of the bath correlation time $\tau_{c,\mathrm{single}}$ at each nitrogen concentration for $^{14}$N (blue) and $^{15}$N (orange), with fits to Eq. (8). Step histograms are the simulated data, smooth curves the fitted inverse Gaussians. The $^{15}$N distribution is displaced to shorter $\tau_c$ at every concentration, by the constant factor set by the resonant-channel ratio of Eq. (6). Insets: the same two fitted distributions magnified about their maxima, with the mode of each marked by a dotted line and the ratio of the two modes given above the arrow. The shift of the maximum is the clearest single-number signature of the isotope effect in these distributions.